\documentclass[preprint,journal]{vgtc}            

\onlineid{0}

\preprinttext{To appear in IEEE Transactions on Visualization and Computer Graphics (VIS 2024).}

\vgtccategory{Research}

\vgtcpapertype{Evaluation}

\title{Talk to the Wall: The Role of Speech Interaction\\in Collaborative Visual Analytics}

\author{%
  \authororcid{Gabriela Molina Le\'{o}n}{0000-0002-9223-2022},
  \authororcid{Anastasia Bezerianos}{0000-0002-7142-2548},
  \authororcid{Olivier Gladin}{0009-0007-5810-6704}, and 
  \authororcid{Petra Isenberg}{0000-0002-2948-6417}
}

\authorfooter{
  \item
  	Gabriela Molina~Le\'{o}n is with the University of Bremen.
  	E-mail: molina@uni-bremen.de.
  \item
  	Anastasia Bezerianos is with the Université Paris-Saclay, CNRS, Inria.
  	E-mail: anastasia.bezerianos@universite-paris-saclay.fr.
  \item
  	Olivier Gladin is with Inria.
  	E-mail: olivier.gladin@inria.fr.

  \item Petra Isenberg is with Université Paris-Saclay, CNRS, Inria.
  	E-mail: petra.isenberg@inria.fr.
}

\abstract{%
  We present the results of an exploratory study on how pairs interact with speech commands and touch gestures on a wall-sized display during a collaborative sensemaking task. Previous work has shown that speech commands, alone or in combination with other input modalities, can support visual data exploration by individuals. However, it is still unknown whether and how speech commands can be used in collaboration, and for what tasks. To answer these questions, we developed a functioning prototype that we used as a technology probe. We conducted an in-depth exploratory study with 10 participant pairs to analyze their interaction choices, the interplay between the input modalities, and their collaboration. While touch was the most used modality, we found that participants preferred speech commands for global operations, used them for distant interaction, and that speech interaction contributed to the awareness of the partner's actions. 
Furthermore, the likelihood of using speech commands during collaboration was related to the personality trait of agreeableness. Regarding collaboration styles, participants interacted with speech equally often whether they were in loosely or closely coupled collaboration. While the partners stood closer to each other during close collaboration, they did not distance themselves to use speech commands. From our findings, we derive and contribute a set of design considerations for collaborative and multimodal interactive data analysis systems.
  All supplemental materials are available at \url{https://osf.io/8gpv2}.
}


\teaser{
  \centering
  \includegraphics[width=\linewidth, alt={Three photos. In photo A, two people are standing in front of the display, and one person is highlighting a sentence on a document by tapping with their finger. In photo B, someone stands close to the display and selects a document by tapping on a radio box on the document header. In photo C, someone is talking on a microphone while looking at some documents on the display.}]{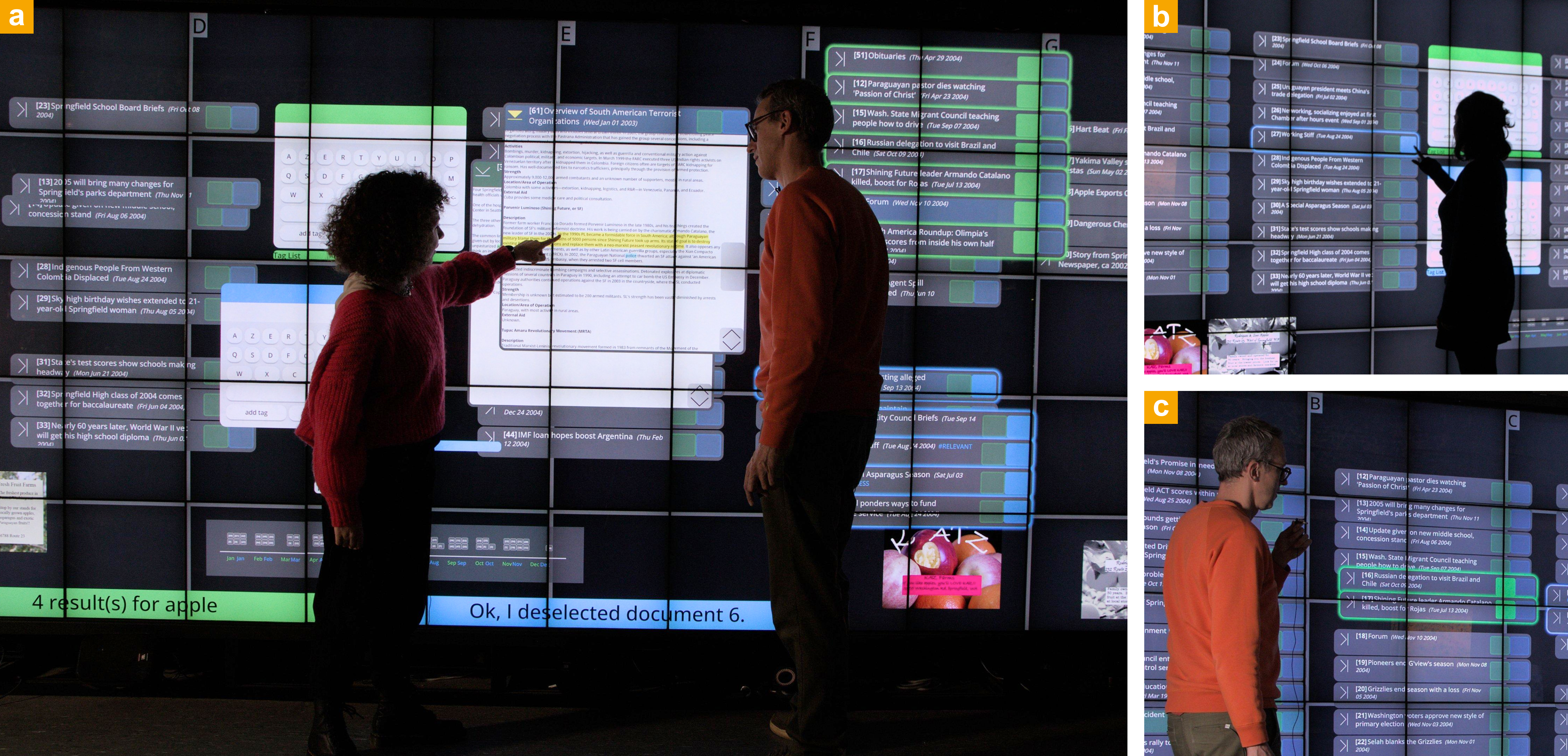}
  \caption{%
  	Collaborative sensemaking system on the wall-sized display: (a) Participants working together on solving the task. (b) Selecting a document via touch interaction. (c) Searching for a keyword via speech interaction.%
  }
  \label{fig:teaser}
}

\graphicspath{{figs/}{figures/}{pictures/}{images/}{./}} 

\usepackage{tabu}                      
\usepackage{booktabs}                  
\usepackage{lipsum}                    
\usepackage{mwe}                       

\usepackage{mathptmx}                  

\PassOptionsToPackage{svgnames}{xcolor}
\usepackage{todonotes} 
\usepackage{siunitx} 
\usepackage {soul} 
\usepackage{layouts} 

\def\oursystem{\textsc{TouchTalkInteractive}} 
\definecolor{speechblue}{RGB}{0, 98, 204}
\newcommand{\speech}[1]{\emph{\textcolor{speechblue}{#1}}}
\newcommand{\newtext}[1]{{\color{red}{#1}\normalfont}} 

\usepackage[skip=5pt]{caption}

\makeatletter 
\renewcommand\st[1]{\@bsphack\@esphack}%
\makeatother

\begin{document}


\firstsection{Introduction}
\maketitle
Wall displays come with a set of benefits and challenges for the visualization and analysis of large datasets. They provide more screen real-estate and offer people ``space to think'' \cite{andrews11}. The large physical size promotes physical navigation for better performance \cite{ball07} and facilitates co-located collaboration \cite{Isenberg:2011:CVD}. As data grows in size and complexity, combining the knowledge and skills of multiple people can facilitate sensemaking tasks \cite{qu08}. 
When two or more people collaborate in front of a wall display, their interactions should not interfere 
with the 
coordination of tasks and activities between team members. 

Interacting with large displays requires techniques that may be unlike the standard mouse and keyboard interactions usually coupled with windows, icons, menus, and pointers (WIMP) interfaces. New technologies now allow us to interact with data using all our physical senses. Ten years ago, already, Roberts et al.~\cite{roberts14} pointed to the importance of future research on how these technologies could be fluidly integrated.
On large displays, touch input is a classic support for close-up interaction via direct manipulation \cite{shneiderman83}. Touch, however, cannot be the only input modality for wall displays because people often stand far away from the wall to look at overviews of the shown data. For interaction distant from the wall, researchers have proposed techniques that do not require touching the display such as laser pointers \cite{myers02}, mobile phones/smartwatches \cite{chapuis14,horak18}, gestural interfaces \cite{nancel11}, and speech~\cite{saktheeswaran20}. Integrating different input modalities is a research challenge in this context because the close-up and distant interaction techniques need to be seamlessly combined 
to support sensemaking activities. In addition, it is not only important how input modalities integrate for one analyst but also how they support multiple analysts working together.

Our work focuses on exploring two specific input modalities during collaborative sensemaking: touch and speech. We chose touch as the most common type of close-up modality for wall displays. Our choice of speech is motivated by the outcome of our recent collaborative elicitation study \cite{molina24}, where participants preferred speech commands over touch, pen, and mid-air gestures to explore data on a large vertical display. Speech (or natural language) interaction is also an interesting candidate to study as it is not yet clear how its use may influence the communication between collaborators, their awareness of each others' activities, and whether personal characteristics may influence the choice of using speech at all. As little is still known about how speech and touch may be used in practice during collaborative analysis, we conducted an exploratory study, focused on how pairs solve a sensemaking task. More specifically, we were interested in their modality choices, their movements in front of the display, and how multimodal interaction and collaboration styles intertwine. 

To investigate these aspects, we designed and implemented \oursystem{}, a wall-based interactive system supporting touch and speech commands for co-located collaborative work. We conducted a study with 20 participants (10 pairs) who solved a fictitious mystery by interacting with a document collection on the wall display. We included interaction techniques facilitated by both modalities and allowed participants to freely choose between them. We analyzed their choices, movements, and collaboration styles and report on our findings regarding how the interaction modalities were used in collaboration. 

In summary, we make the following contributions:
\begin{itemize}
    \item We present insights on the interaction choices, movements, and collaboration styles of the participants and the interplay with speech interaction.
    \item We examine how personality traits relate to the use of input modalities, such as speech commands.
    \item We derive a set of design considerations on how to support co-located collaborative work leveraging speech interaction on wall-based 
    interactive systems.
    \item We release \oursystem\ as an open-source tool to support reproducibility.
\end{itemize}


\section{Related Work}
We present relevant work related to our research on the topics of interaction \st{on} with large vertical displays, natural language interfaces, collaborative visual analytics, and personality traits.

\subsection{Interaction with Large Vertical Displays}

Large vertical displays provide more pixels and more space to interact with \cite{belkacem22}. 
As the space in front of the display allows physical navigation, which can lead to better performance \cite{ball07}, wall-based systems should provide different ways of interacting from varying distances, to make the most of the surface and space.
Touch is the standard input modality to support direct manipulation, but it comes with the disadvantage that the display areas beyond the arm's reach become inaccessible. Thus, researchers have come up with interaction techniques to virtually move the display content towards the user \cite{riehmann20}. Interacting with a pen or stylus is another type of direct manipulation that people leverage to write annotations \cite{molina24} and author visualizations \cite{walny12}. While touch and pen can be a powerful combination \cite{hinckley10}, we decided to leverage touch only as it is the most common modality supported and it requires no additional hardware. 
However, direct manipulation requires participants to stand close to the wall display. Thus, we need other input modalities to facilitate interaction from a distance. For instance, Nancel et al.~\cite{nancel11} evaluated and proposed mid-air gestures to navigate through pan and zoom interactions on wall displays. Another possibility is to use additional devices, such as mobile phones \cite{langner19} and smartwatches \cite{horak18} to interact from a distance. However, our work focuses on supporting distant interaction without needing additional handheld devices.

Previous work has studied the combination of multiple input modalities on large vertical displays. Srinivasan and Stasko \cite{srinivasan18} proposed combining direct manipulation and speech commands to interact with network visualizations and found that participants chose different modalities across tasks, e.g., speech for filtering and touch for highlighting connections between nodes. DataBreeze \cite{srinivasan21} combines touch, pen, and speech to interact with unit visualizations. While interacting, participants tended to use speech for global actions and touch-based context menus for local actions. Accordingly, we incorporated examples of such interactions in our system, to examine whether the same happens in the context of collaborative work.
Saktheeswaran et al.~\cite{saktheeswaran20} found that when comparing touch, speech, and touch-and-speech interactions, participants preferred the multimodal interactions. However, Molina Le\'{o}n et al.~\cite{molina24} found that people preferred speech commands, over touch, pen, and mid-air gestures, alone or in combination, for a series of exploratory tasks. Therefore, we examine the use of touch and speech in a real-world system to study how participants choose to interact.

\subsection{Natural Language Interfaces}
Natural language interfaces (NLI) facilitate interaction through written or spoken language. They allow users to express their intent without having to translate questions into data attributes \cite{grammel10} but suffer from discoverability issues and depend on feedback \cite{srinivasan17}.
Eviza \cite{setlur16} is an example of such an interface, supporting visual data analysis through written text, and giving the user a sense of dialogue by supporting follow-up queries. 
Gao et al.~\cite{gao15} designed DataTone, which supports both written and spoken text and addresses the issue of language ambiguity by generating \st{ambiguity} widgets for users to refine their queries. 
Recently, Guo et al.~\cite{guo24} proposed a deep-learning solution that generates visualizations to answer multiple aspects considered relevant to a question posed \st{provided} by the user. 
While these solutions are all desktop-based, \st{other researchers have designed} there are also NLIs designed for other devices. 
Kassel and Rohs \cite{kassel18} designed Valetto, a tablet-based interface that supports visualization authoring through speech and touch. The authors assigned each modality to specific actions, but their evaluation revealed that redundancy would be preferable. Tabalba et al.~\cite{tabalba23} designed an NLI that allows pairs to author charts on a large display, using their group conversation as additional query context.
In contrast, we study interaction techniques facilitated through both touch and speech. While we do not focus on conversation or query refinement, we seek to support diverse low-level actions to give the option of choosing between two modalities.

A common challenge to facilitate natural language interaction is the technological capabilities to recognize natural language. 
Accordingly, Narechania et al.~\cite{narechania21} created a toolkit to author visualizations based on a written query and related systems made use of HTML5 speech recognition APIs to translate from spoken language to text \cite{srinivasan20inchorus}. To circumvent common accuracy issues, we trained a deep-learning model with the vocabulary of our dataset and integrated it into our system.

\subsection{Collaborative Visual Analytics}
Researchers have proposed different ways of supporting multi-user interaction on large displays. Langner et al.~\cite{langner19} proposed using touch and touch-enabled mobile devices to allow participants to interact from different distances from the display. Badam et al.~\cite{badam16} designed interaction techniques combining proxemics with mid-air gestures to interact with lenses on wall displays. James et al.~\cite{james23} proposed to leverage augmented reality to extend the interaction possibilities in the space in front of the display. We designed a system to leverage touch and speech interaction between two users in parallel if they want.

When people do not only interact with a system, but also with each other, social aspects need to be considered, such as collaborative distance, awareness, and communication among team members \cite{lee12}. We examine how the use of speech may influence these aspects, e.g., encouraging people to move away from each other before using a speech command. We studied how speech interaction may influence awareness and its interplay with human communication. 
Territoriality and privacy are other aspects to consider. For example, Reipschläger et al. \cite{reipschlager21} proposed using augmented reality to extend the display surface by creating a personal virtual space for each collaborator. While these aspects are out of the scope of our work, we designed our system to provide a set of interface elements (e.g., tag list, virtual keyboard) that would allow participants to work alone if desired.

Regarding the use of multiple input modalities for collaborative work, Tse et al.~\cite{tse08} explored how combining speech and hand gestures affected collaborative work on tabletops. The authors identified speech use in parallel commands as a design issue, i.e., if speech commands are used often for actions that people need to perform in parallel, one person may decide to work sequentially to avoid voice overlapping. Their findings inspired us to study the effects of speech commands on wall displays.
Furthermore, the way people work together can qualitatively vary depending on how closely or loosely they collaborate. Tang et al.~\cite{tang06} proposed a set of codes to classify collaboration styles according to how strong the relationship between the activities of each person was. Isenberg et al.~\cite{isenberg10} extended the set for analyzing collaborative work on tabletops to take multiple data views into account. Other researchers have extended or adapted the \st{collaboration} styles proposed by Tang et al.\ and Isenberg et al.\ for scenarios with more than one device \cite{brudy18} or for hybrid collaborations \cite{neumayr18}. However, those characteristics do not apply to our scenario, thus, we use the code set of Isenberg et al.\ to analyze the collaborative work of our participants.
We selected the dataset and designed our experiment based on their work and follow-up studies \cite{jakobsen14, andrews10} to facilitate comparison and validation of our findings.


\subsection{Personality Traits and Visualization}
\label{sec:rw-personality}

Designing one-size-fits-all interfaces for visualization systems is problematic because the individual differences among people can affect how a person uses a visualization~\cite{liu20}. Thus, researchers have looked at personality traits to analyze whether and how they may affect the experience of using visualizations. The most known personality research model is the Five-Factor Model. 
As its name suggests, it refers to five personality traits that describe individual differences: extraversion, openness, agreeableness, conscientiousness, and neuroticism.
We focus on the first three traits, as we consider them relevant for speech interaction and collaborative work. \textit{Extroverted} people are outgoing and sociable \cite{volkel20}. In a study on task performance, Green et al.~\cite{green10} found that extroverts solved search tasks faster than introverts. However, introverts gained more insights. Ziemkiewicz et al.~\cite{ziemkiewicz13} found that introverts took more time to analyze the task and, thus, solved more tasks accurately than introverts. We expect that extroverted participants will use speech interaction more often than introverts because they may feel more comfortable speaking publicly.

\emph{Openness to experience} is the second trait that could influence the likelihood of using speech interaction, as it refers to curiosity and the wish to seek new experiences \cite{deyoung12, volkel20}. 
\st{According to Liu et al.~[32], openness is a personality trait under-explored in data visualization research.}
We expect the willingness of the participants to experiment with less common \st{input} modalities (such as speech) may influence their choices.
\emph{Agreeableness} is the third trait we examine, as it refers to being cooperative~\cite{volkel20} and is considered a positive trait for people to succeed in collaborative work~\cite{liu20}. We expect that the use of speech interaction negatively correlates with agreeableness. In other words, we suspect that participants will be reluctant to use speech so as not to disturb their colleagues.


\section{Experiment Design}
To explore the synergy between speech and touch interaction, we designed an in-person experiment where pairs of participants had to solve a sensemaking task collaboratively. Our goal was to investigate the following research questions:

\begin{description}
    \item[RQ1] How do partners use touch and speech individually while working as part of a team? 
    \item[RQ2] How are touch and speech used during collaboration? When is each modality chosen and how does it relate to in-team communication and coordination?
\end{description}

We asked each pair to solve a fictitious mystery described in a collection of text documents that could be interacted with on a touch and speech-enabled wall display. We designed the experiment inspired by two prior collaborative studies on wall displays \cite{isenberg10,jakobsen14}. Following a prior VAST challenge \cite{grinstein06}, the researchers asked participants to find a weapons-smuggling plot hidden among a large number of text documents. Teamwork can be helpful to solve the task as more documents can be read together and participants can discuss their hypotheses. In our system, participants were free to use any supported touch gestures and speech commands. To understand participants' choice of modalities, we ensured that all commands were available in both speech and touch; see \cref{sec:design} for details on the interaction techniques. We video-recorded the sessions, tracked the position of the participants, and logged their interactions with the system.

Before the main experiment, we conducted two pilot studies. The first helped us finalize the list of interaction techniques the system should support to solve the task. The second pilot study with another participant pair helped us identify usability issues that \newtext{we} then solved before the main study, and to refine the protocol and the setup regarding audio capture when multiple people were talking simultaneously.

\subsection{Dataset and Task}
\label{sec:dataset}
We created an adapted version of the dataset used in the ``Stegosaurus'' scenario from the interactive session of the 2006 VAST challenge \cite{grinstein06}. 
Our dataset contained 61 text files and four images. 
The text files were news articles from a fictitious town called Springfield\footnote{We modified the town name (previously named Alderwood) and other proper nouns in the dataset to ensure the speech engine would understand them.} and other miscellaneous documents, such as a list of diseases and a list of terrorist organizations. 
From the 61 text files, 10 news articles contained relevant information about a hidden weapon-smuggling plot, seven provided background information, and the miscellaneous documents gave some important clues about the evidence (e.g., the characteristics of a chemical involved). 
Similar to previous work~\cite{isenberg10, jakobsen14}, the pairs had one main task: generating hypotheses about what happened.
The task encourages participants to collaborate loosely and closely on different parts of the task \cite{jakobsen14}  (e.g., reading documents and discussing hypotheses), which allowed us to study various modes of collaboration.

\subsection{Personality Traits}
 We measured three personality traits per participant: extraversion, openness to experience, and agreeableness (see \cref{sec:rw-personality}). We hypothesized that participants with higher extraversion and openness to experience would be more at ease to use speech commands, as they may feel more comfortable speaking publicly and using new technology; whereas participants with high agreeableness may be reluctant to use speech so as to not distract others. We decided to examine personality traits because, in the first pilot study, both participants suggested that personality was an important factor regarding the choice to use speech interaction. For instance, one participant indicated that \textit{``I found that I do not want to use speech in front of him (...) I am just an introverted person.''}

\begin{figure*}[t]
  \centering
  \includegraphics[width=\linewidth, alt={A screenshot of the interactive system showing the dataset. The interface includes two keyboards, two tag lists, a timeline, two message bars, and the images and documents from the dataset. The documents are organized in a grid layout and three of them are open. A few sentences are highlighted on those documents.}]{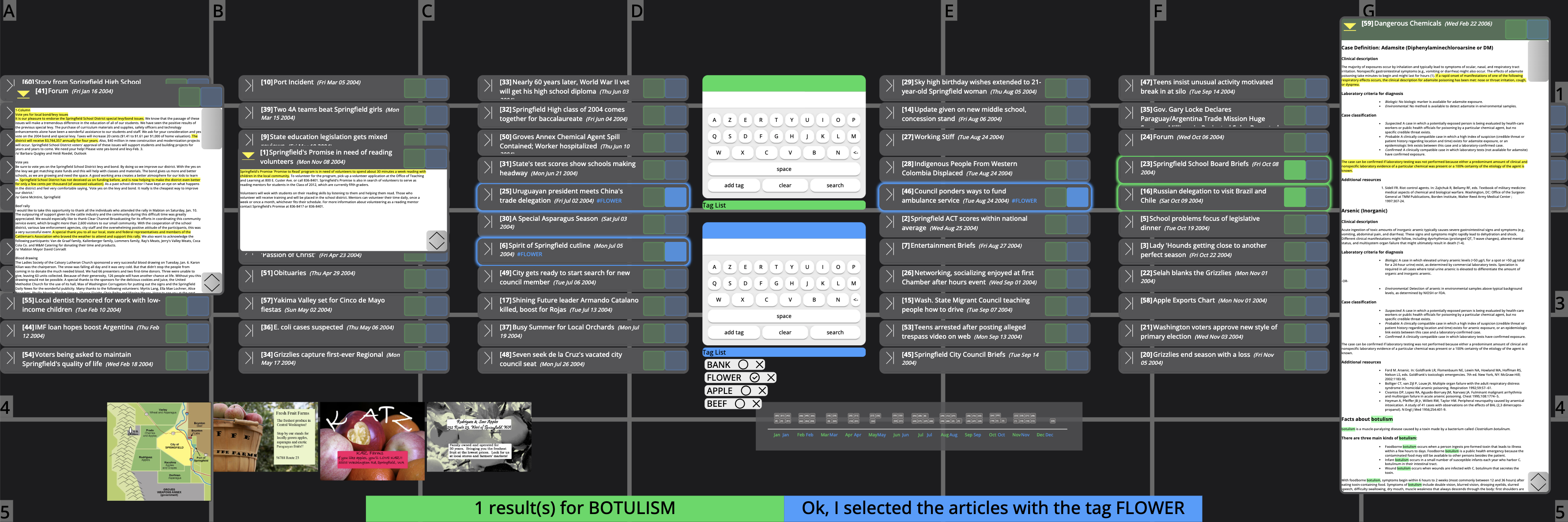}
  \caption{Screenshot of \oursystem. The blue and green colors identify the interface elements associated with each participant. The bars on the bottom of the screen give feedback to the user regarding their last speech command and any search-related tasks with either touch or speech.}
  \label{fig:screenshot}
\end{figure*}

We used the IPIP-NEO-60 questionnaire by Maples-Keller et al.~\cite{mapleskeller19} of the NEO PI-R domains from the instrument of Costa and McCrae~\cite{costa92}. Based on the answers, we calculated three scores per participant that we could operationalize to describe each trait and compare them to the total of speech commands. 
While dozens of personality assessment instruments exist, many are copyrighted and paid-only. Therefore, we chose the freely available IPIP–NEO–60, which is easily accessible for future replications. It includes 12 questions per trait, and it is, thus, more practical than others available with over 300 questions.

\subsection{Procedure}
We started the experiment by explaining our research motivation and handing participants a consent form to read and sign. If both participants gave consent, we gave each of them a pre-questionnaire to fill out. The questionnaire included five questions about their previous experience with interactive technologies, two questions about their collaboration experience with their partner, two demographic questions, and 36 questions from the personality traits questionnaire. 
Afterward, we proceeded to a 10-minute training session. The experimenter introduced the system and then participants went through a list of all supported interactions and performed every interaction using speech and touch gestures at least once. We proceeded to the main phase of the study once participants felt confident enough to interact with the system independently. 
We also provided participants with a cheat sheet (one page) of all interaction techniques supported by the system, describing the necessary actions for using them with touch and speech. Most of them kept the sheet in their hand while solving the task.

The main phase of the study took 45 minutes. Participants started by reading a one-page background document introducing the task scenario. This document told the story of a mysterious incident in the fictional city of Springfield that the police wished to investigate further. 
To generate hypotheses, explore the dataset, and solve the mystery participants could interact with the system as they wished. Participants could ask questions to the experimenter about the system, but they did not receive additional help to solve the mystery. 
Once the 45 minutes were over, participants filled out a post-questionnaire about their experience in terms of collaboration and input modalities. Afterward, we conducted a short interview where we asked them to elaborate on what they found out about the mystery and their impressions of the input modalities, as well as of the collaborative work with their partner. 
We did not control the initial position of the participants, they were free to move in the space in front of the wall display as they wished. We also did not measure their performance because our goal was to capture holistic, qualitative, observational data to understand how they interacted with each other and the system---and not to compare accuracy or speed that in this task is highly dependent on text reading speed and understanding. 

\subsection{Data Collection and Analysis}
To answer \textbf{RQ1}, we logged interactions to know how often each participant used each input modality and for which actions.  
To measure the distance to the screen, we used a Vicon motion tracking system to document the movements of the participants in front of the wall display.

We video-recorded the main study phase with a camera positioned at the back of the room and a microphone above the display. We used two wireless microphones to perform the speech commands. A speech-to-intent engine translated the commands directly into machine-readable text in JSON format, including only the
information necessary to process the command.
We analyzed the videos qualitatively to identify the interaction and collaboration styles of the participants to help us answer \textbf{RQ2}. 
One author of this paper coded the videos in two passes with the help of a commercial coding tool. The first pass was meant to validate and complete the interaction logs and the second was focused on coding the collaboration styles. We applied the codes associated with collaboration styles proposed by Isenberg et al.~\cite{isenberg10}, except for their SIDV code that refers to multiple copies of the same document, as our system showed each document only once.

The post-questionnaire and the interview helped to obtain subjective feedback about the collaborative experience and the interactions. We generated the first transcription of the interviews with an automatic tool that we then revised manually. We analyzed the interviews through an open coding scheme based on grounded theory.

\begin{table*}[t]
  \caption{Interaction techniques supported by the system. Each interaction can be performed with touch gestures and speech commands.}
  \label{tab:techniques}
  \scriptsize%
  \begin{tabular}{lp{76mm}p{60mm}}
    \toprule
    Action & Touch gesture \includegraphics[height=4mm]{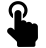} & Speech command ~\includegraphics[height=4mm]{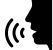}\\
    \midrule
    Open document & Tap on the left corner of header (toggle). & \textit{``Open document} \speech{12}\textit{''} \\
    Close document & Tap on the left corner of header (toggle). & \textit{``Close document} \speech{12}\textit{''}\\
    Select document & Tap on radio box on the document header. & \textit{``Select document} \speech{12}\textit{''}\\
    Select documents per month & Tap on the month name on the timeline. & \textit{``Select documents of the month} \speech{May}\textit{''}\\
    Select documents per tag & On the tag list, tap on the tag name & \textit{``Select documents with the tag} \speech{money}\textit{''}\\
    Deselect document & Tap on the header of the selected document & \textit{``Deselect} \speech{12}\textit{''}\\
    Deselect all & Long press on canvas to open the context menu. Tap ``Deselect.'' & \textit{``Deselect all''}\\
    Sort & Long press on canvas to open the context menu. Tap ``Sort.'' Then, tap on sort parameter. & \textit{``Sort by} \speech{date}\textit{''}\\
    Resize document & Drag bottom-right corner of an opened document. & \textit{``Make document} \speech{12} \textit{smaller''}\\
    Move a document & Drag document by its header. & \textit{``Move document} \speech{12} \textit{to column} \speech{A}, \textit{row} \speech{3}\textit{''}\\
    Move multiple documents & Drag selected documents by dragging one of them. & \textit{``Move selected to column} \speech{A}, \textit{row} \speech{3}\textit{''}\\
    Navigate in document & Drag scrolling bar inside document. & \textit{``Scroll} \speech{down} \textit{in document} \speech{12}\textit{''}\\
    Mark document & Tap on sentence (toggle). & \textit{``Mark [sentence} \speech{1} \textit{in] document} \speech{12}\textit{''} (first by default)\\
    Clear sentence & Tap on sentence (toggle). & \textit{``Clear sentence} \speech{1} \textit{in document} \speech{12}\textit{''}\\
    Clear document & Tap on sentences (toggle). & \textit{``Clear document} \speech{12}\textit{''}\\
    Search & Tap phrase on keyboard. Then, tap on ``search.'' & \textit{``Search for} \speech{flowers}\textit{''}\\
    Clear search & Tap on ``clear'' on keyboard. & \textit{``Clear search''}\\
    Add tag & Type tag name on keyboard. Then, tap on ``add tag.'' & \textit{``Add the tag} \speech{flowers}\textit{''}\\
    Assign tag to document & Select document. Tap on O icon next to the tag. & \textit{``Assign tag} \speech{flowers} \textit{to the selected documents''}\\
    Remove tag from document & Select tagged document. Tap on O icon next to the tag. & \textit{``Remove tag} \speech{flowers} \textit{from the selected documents''}\\
    Delete tag & On the tag list, tap on X icon next to the tag. & \textit{``Delete the tag} \speech{flowers}\textit{''}\\
    \bottomrule
  \end{tabular}
\end{table*}

\section{System and interaction design}
\label{sec:design}

We designed and implemented a visual analytics system called \oursystem\ to support co-located collaborative sensemaking on wall-sized displays, using touch and speech interaction. Its functionality is inspired by previous research-oriented systems created to support collaborative tasks \cite{isenberg09, jakobsen14, andrews10} and its speech support by studies on interaction with large vertical displays \cite{saktheeswaran20, molina24}. While prior wall display systems support speech and other modalities for a single user \cite{saktheeswaran20, srinivasan21} or speech for visualization authoring in a group context \cite{tabalba23}, ours enables the use of speech interaction for collaborative sensemaking. 
Our goal was to use the system as a \emph{technology probe}~\cite{Hutchinson2003} with two goals: a) to help understand how well the proposed solution addresses interaction and collaboration needs and b) to inspire users and researchers to reflect on speech interaction in collaboration.  

Our prototype is designed to assist two people in exploring and making sense of a large collection of documents through touch gestures and speech commands. Each document is visually represented as a unit, that can be opened or extended to read the content. The system includes a timeline to visualize and select the documents according to the temporal dimension and supports images. By default, the images are positioned at the bottom of the screen, as previous work suggests that the lower area of wall displays should not be used for data representations \cite{bezerianos12}. 
Each person can interact with one or multiple documents at a time, and both persons can interact simultaneously. 
Speech recognition is activated with a click on a wireless mouse. Thus, using a wake word, such as ``Hey Siri'', is not required, and participants can be confident that the system only listens to them when they wish to.
Each person has a designated color (blue or green) and a set of interaction widgets in their color, e.g., a keyboard to search for keywords. That way, participants are free to work individually if they wish to.

\subsection{Interaction Techniques}
Our goal was to support a comprehensive list of interactions with the documents, both through touch gestures and speech commands, so that each of the two participants could interact in parallel while choosing any modality. Although single multimodal interactions would have been possible (i.e., combing speech and touch in one command), we kept the modalities separate to compare our results to the elicitation study of our related work~\cite{molina24}, and to examine when and in what contexts participants would choose either modality.
We present the 20 interaction techniques the system supports in \cref{tab:techniques} (for the speech commands, we show only the main examples because synonyms were supported). We carefully chose them to facilitate co-located and synchronous work, based on prior work on multimodal and collaborative interaction \cite{isenberg10,jakobsen14,molina24,saktheeswaran20}.
When the system recognized a speech command, it provided feedback by showing a message on the corresponding bar at the bottom of the screen (e.g., ``Ok, I selected document 6''). If the recognition failed, the message ``Sorry, I did not understand that'' appeared.

\subsection{Setup}
Our wall display was 5.91 × 1.96 meters in size, composed of 75 liquid-crystal displays (LCDs) with a total resolution of 14400 × 4800 pixels (at 60 ppi), controlled by a cluster of 10 computers.
The wall was placed in a room of \SI{31.5}{\metre\squared}, 
and it recognized touch input via infrared light. For recording the position of the participants, each of them wore a headband tracked by a VICON motion tracking system.
Each participant had a wireless mouse and a microphone. They could activate the speech recognition by clicking on the mouse.

\oursystem\ is a web-based application, implemented in Typescript and Svelte. The documents are provided to the system as JSON files.  
We implemented the speech recognition on Rhino, a deep learning speech-to-intent engine by Picovoice~\cite{picovoice}. We trained an English-based model according to the vocabulary used in the dataset documents to enable custom speech commands, with a reported accuracy above other commercial solutions \cite{picovoice_benchmark}. The open-source code is available at \href{https://gitlab.inria.fr/aviz/TouchTalkInteractive}{https://gitlab.inria.fr/aviz/TouchTalkInteractive}.

\section{Results}
We recruited 20 participants (five identified as women, and fifteen as men) distributed into 10 pairs. We recruited the pairs from within a research organization, as the goal was to recruit people who would interact with data as part of their work. Participants were required to be at least 18 years old and fluent English speakers. The study was approved by the corresponding IRB (Inria COERLE, opinion 2023-39).
Participants worked or studied at the institution, but had diverse backgrounds. We had participants from Europe, the Americas, and Asia. However, cultural background was not a factor we controlled.

The average age was 27 years. All participants reported interacting with touch-based systems daily. Nine persons used speech interaction at least monthly, while four had never used it before. For 14 participants, this was the first time interacting with a wall display. 
As the task was similar to solving a puzzle, we asked how often participants played such games. Eighteen participants had played at least once, seven did so monthly or daily. 
We did not require that participants knew each other before the experiment to get a diverse sample of collaborators. Six pairs were familiar with each other before participating and four were not. 
Sixteen participants reported to work closely with others at least once a week.
All pairs were able to find important clues about the mystery. At the end of the interaction phase, several groups expressed that they would have liked to have more time with the task. This suggests that people engaged with the game-like setting.

\subsection{Input Modalities per Action}
In total, we registered $4,020$ interactions from the 20 participants. Of those, $3,633$ were touch gestures and 387 were speech commands. Given that some actions happened more often than others (e.g., moving documents), we analyzed the interactions per action  \textbf{RQ1}.
The system supported 20 actions. However, we excluded interactions for the action \emph{select documents per month} from our analysis (77 interactions), as we had technical problems with that feature in the first two sessions.

Participants were free to choose how to interact to solve the sensemaking task, after having tried all the interaction techniques with both modalities in the training session.
Overall, participants made use of all the interaction techniques provided. The most used ones were searching for keywords, opening, closing, and moving documents, as well as highlighting and clearing sentences of interest in the documents. Most groups created and assigned tags to documents, but only a few removed or deleted those tags. Participants selected documents one by one more often than per tag or month. Similarly, they deselected individually more often than \emph{deselecting all}. 

\begin{figure*}[tbp]
  \centering
  \begin{subfigure}[b]{0.5\linewidth}
  	\centering
  	\includegraphics[width=\textwidth, alt={Stacked bar chart showing the percentages of modality use per action on the X axis, with 20 actions on the Y axis. The input modality is encoded as the color of the bars. The actions sort, deselect all, and clear document are the only actions with speech use covering more than 50\% of the interactions.}]{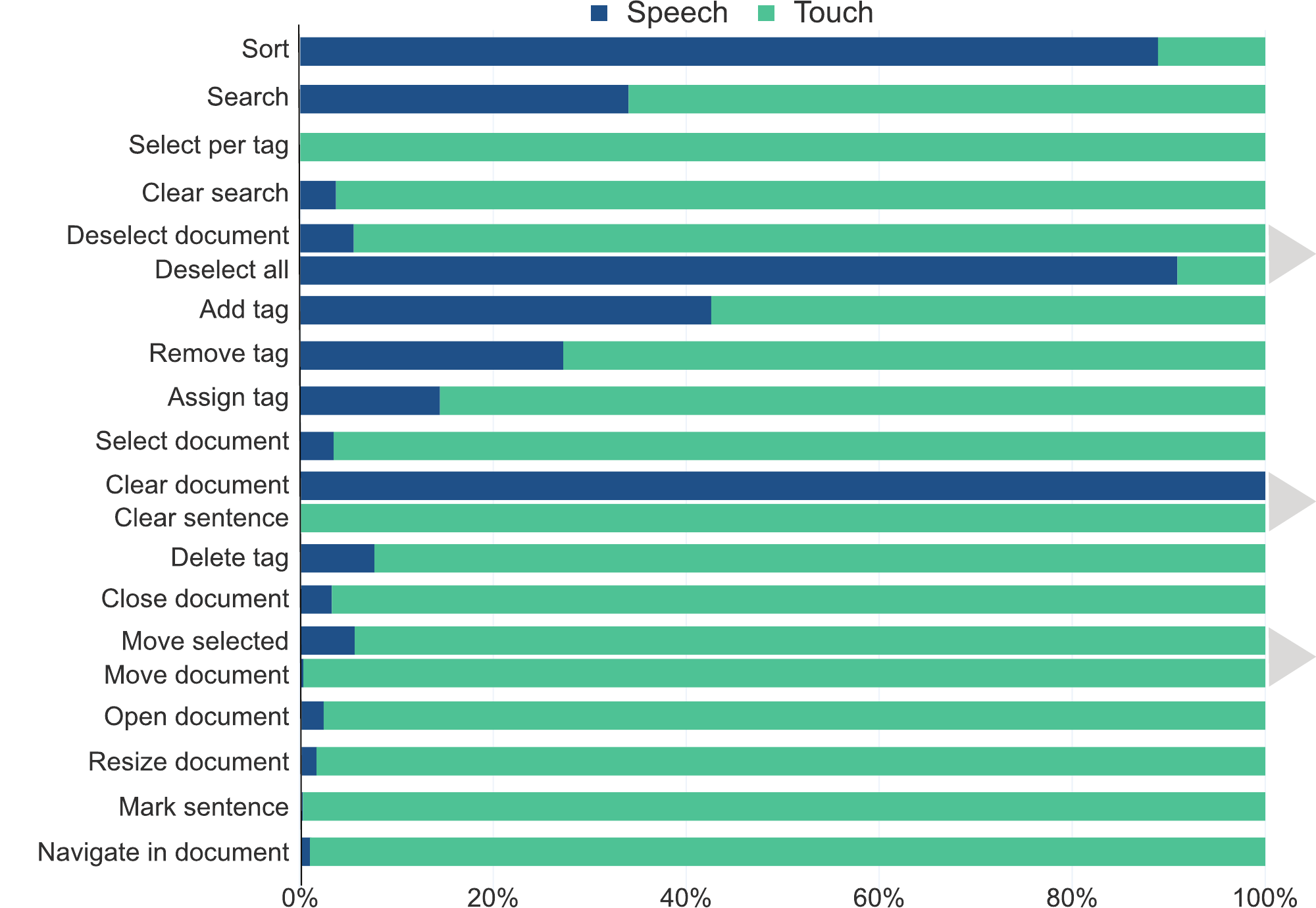}
  	\caption{Distribution of the interactions across actions and modalities (\% of interactions).}
  	\label{fig:pertask}
  \end{subfigure}%
  \hfill%
  \begin{subfigure}[b]{0.5\linewidth}
  	\centering
  	\includegraphics[width=\textwidth, alt={Stacked bar chart showing the modality preferences of the participants per action category. There are 17 categories on the Y axis and the participant count (maximum 20) on the X axis. Three action categories correspond to two actions. For example, the category Deselect includes deselecting a single document and deselecting all documents. The categories sort, search, select per tag, and clear search are the only four where touch was preferred by less than half of the participants.}]{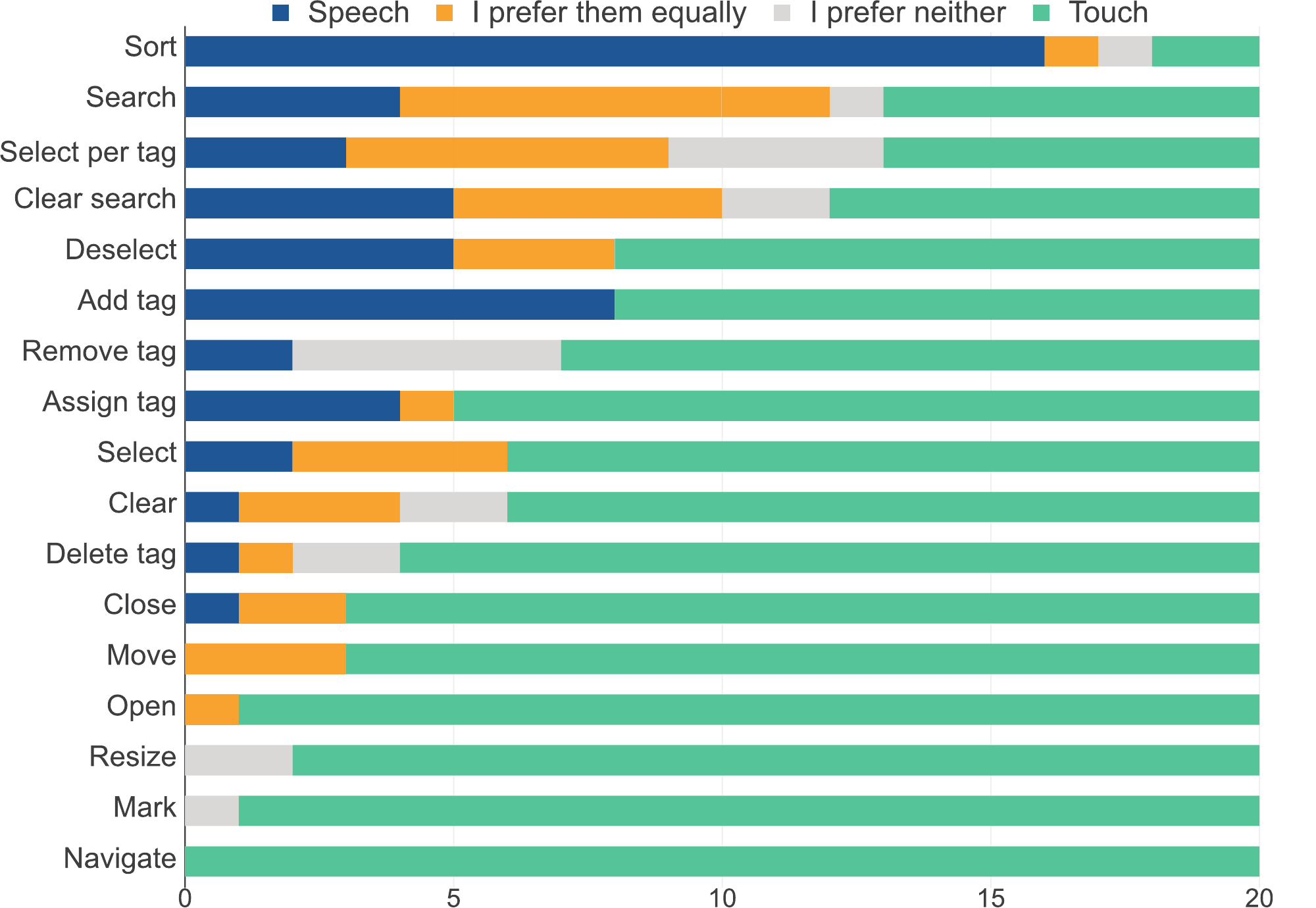}
  	\caption{Input modality preferences of the participants per action category (number of participants).}
  	\label{fig:preference}
  \end{subfigure}%
  \vspace{0.5mm}
  \subfigsCaption{Use and preferences of input modalities per action.}
  \label{fig:ex_subfigs}
\end{figure*}

In \cref{fig:pertask}, we present the recorded interactions distributed across actions, with the proportion of modalities used. We show only the successful interactions (e.g., without speech recognition errors) where the system interpreted the action correctly. Participants mainly used touch interaction, but there were exceptions among the actions. Participants sorted the documents, cleared highlighted documents, and used \emph{deselect all} more often with speech interaction. and two participants used touch (once in both cases).
To deselect all, seven participants used only speech, and only one person used touch.
To search for keywords, 18 participants used the touch-based keyboard, while six used speech commands more often than touch.
To open and close documents, all except one participant used touch gestures more often than speech.
To select single documents, one person used only speech while everyone else used touch more often (or only).
The diversity of the choices suggests that participants were able to use both modalities, but made different personal choices. For instance, while one participant (P12) interacted with speech commands throughout the whole session, another participant (P20) only used speech once.

In \cref{fig:preference}, we present the modality preferences of our participants after interacting with the system. Most participants preferred speech interaction to sort the documents. For keyword search, more than half of the participants preferred speech or favored touch and speech equally. To clear search results, half of the participants preferred speech or both. 
For the remaining actions, touch was preferred. To select documents with a given tag, seven participants preferred touch, but six wished to do it with either modality.
Some participants made remarks about their reasoning when they chose the \emph{I prefer them equally} answer. For example, P2 pointed out that choosing between touch and speech for closing a document depended on their distance from the document. Similarly, for moving documents, the choice would depend on the distance between the current and the desired position of the document on the wall display. Another participant also referred to the distance to the wall and to the document as a decisive factor in choosing how to open a document. For two participants, speech would be the first choice to search for a keyword, and if there would be a recognition error, they would switch to the touch-based keyboard. Another participant considered that searching by touch took too much time.

\begin{figure}[b]
  \centering
  \includegraphics[width=\linewidth, alt={Three plots in a row, each comparing a personality score on the y-axis against the total of speech commands on the x-axis. The plots of agreeableness and openness indicate the possibility of a negative correlation.}]{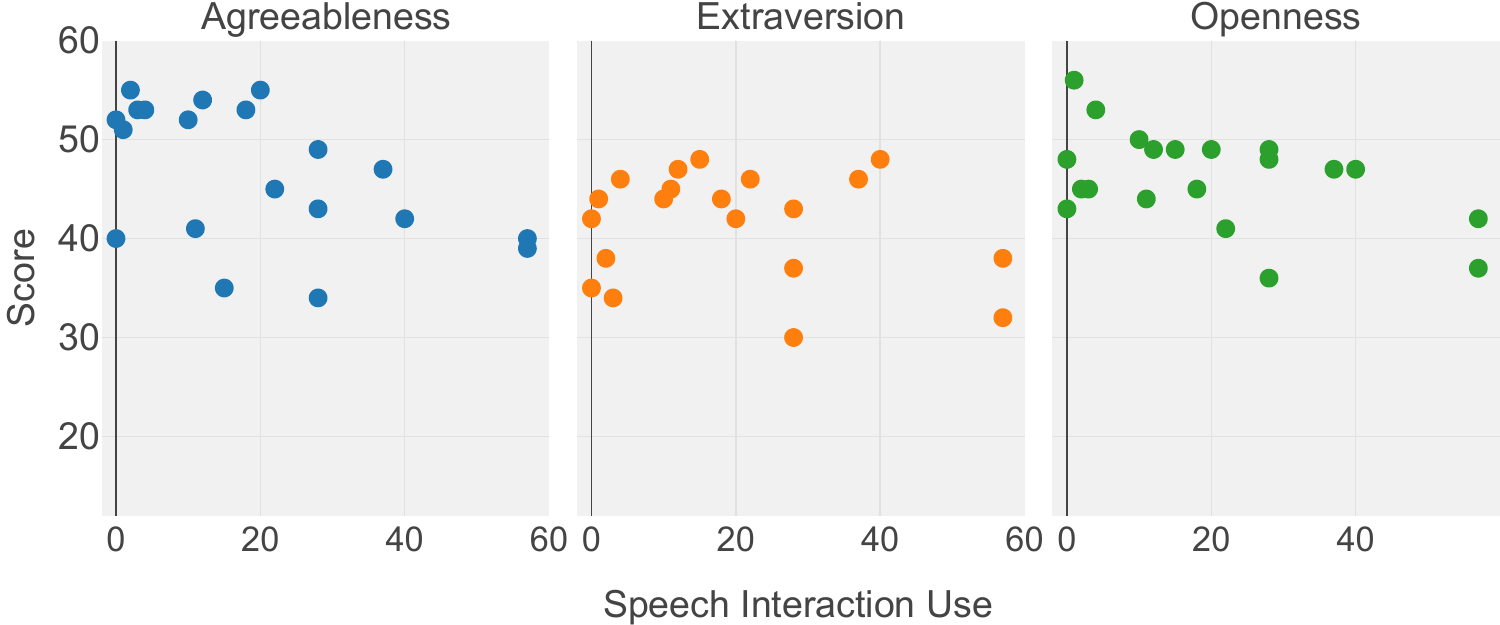}
  \caption{Scatterplots visualizing the scores from each personality trait in relation to the total count of speech commands per participant.}
  \label{fig:agreeableness}
\end{figure}

\subsection{Personality Traits}
Based on their answers to the personality questions in the pre-questionnaire, we divided participants into three groups per trait: those with low, high, or average scores, in comparison to the sample mean and standard deviation, as suggested in the literature~\cite{ipipwebsite, ziemkiewicz13}.
Seven participants scored high on extraversion, six were average, and seven scored low.
Seven participants scored high on openness to new experiences, eight were average, and five were low.
Nine participants scored high on agreeableness, three were average, and eight scored low.
We compared the scores of the three personality traits against the number of times each participant attempted to use speech interaction (regardless of the action and outcome), to examine whether those traits could relate to the tendency of using speech interaction. 
We visualize the relations in \cref{fig:agreeableness}.
We found a significant relation between agreeableness and speech using Spearman's correlation ($\rho = -0.445, p = .0495$), suggesting a moderately strong relationship between how agreeable a participant was and how often they used speech interaction. The more cooperative a person was, the fewer speech commands they used, potentially to avoid interrupting their partner, given that eight participants reported being afraid to distract their partner in the post-questionnaire.
In contrast, the correlation between extraversion and speech use was not significant ($\rho = 0.016, p = .947$). 
Furthermore, contrary to our expectations, speech use and openness to experience correlated negatively, although not significantly ($\rho = -0.351, p = .129$). 

\subsection{Speech Interaction from a Distance}
\begin{figure}[b]
  \centering
  \includegraphics[width=\linewidth, alt={Two heatmaps showing how many touch gestures and speech commands were executed from a specific position in 2D space (up to four meters from the display). While touch gestures clutter in the side regions at less than one meter from the display, the speech commands clutter towards the right region around the one-meter distance.}]{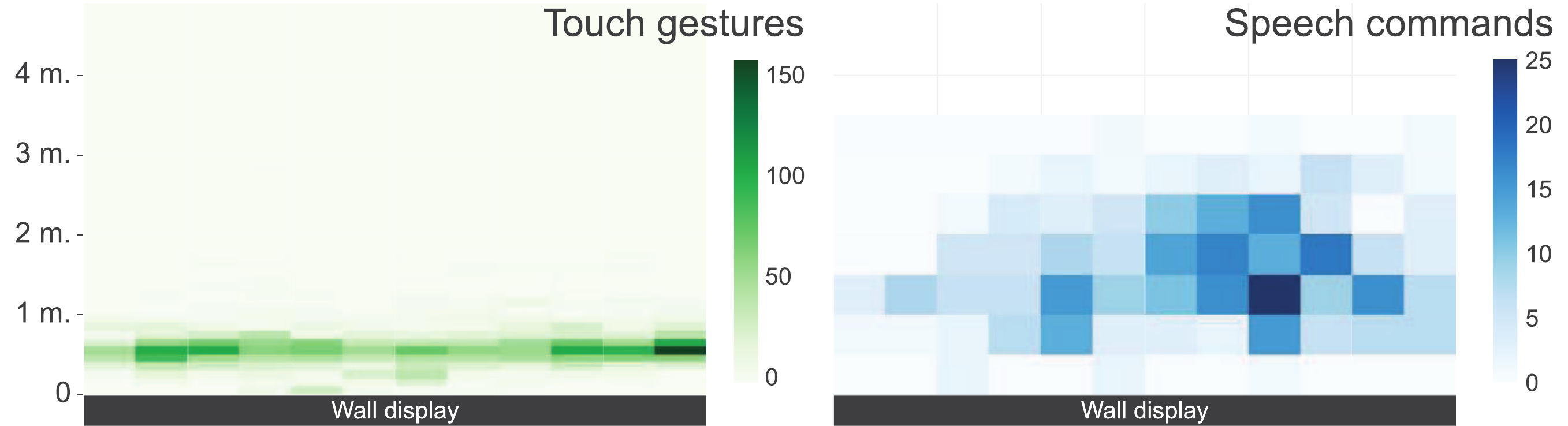}
  \caption{Distribution of the participant positions in the space in front of the wall display while interacting, separated per modality.}
  \label{fig:distances}
\end{figure}

We compared the frequency of speech commands with the distance of the participants from the wall display, based on the interaction logs and the tracking data we collected during the study. We present the distribution of the participant positions per modality in \cref{fig:distances}.

While touch interaction required standing at a close distance, participants could interact with speech from any position. The data we collected about the participants' movements suggest they tended to stay away from the wall while using speech commands. On average, they stood at $1.52$ meters from the display when interacting with speech and at $0.59$ meters when interacting with touch. Furthermore, when a participant used a speech command, the average distance to their partner was $1.88$ meters ($std=0.86$) and the median was $1.80$ m. In contrast, when someone interacted with touch gestures, the distance between partners varied more, with an average of $1.94$ meters ($std=1.25$) and a median of $1.52$. While the larger variation of the distances associated with touch may be due to the higher frequency of touch gestures overall, the distances indicate that participants did not intentionally go away from their partner to use speech commands.

\subsection{Collaboration Styles}
To answer \textbf{RQ2}, we qualitatively coded the collaboration styles of the participants throughout the sessions according to the codes proposed by Isenberg et al.~\cite{isenberg10}. Overall, participants spent more time in closely coupled collaboration than in loosely coupled collaboration. On average, the groups spent 54\% of the session collaborating closely and 36\% loosely. The remaining 10\% of the time corresponds to periods of no clear collaboration, e.g., when participants talked to the experimenter.All pairs reported to have divided tasks among the two persons while solving the mystery. In the post-questionnaire, we asked each participant to estimate the proportion of time they spent working together with their partner, alone researching a shared question, and alone researching their own question. On average, participants reported to have spent 31\% of the time working with their partner ($std = 21.05$), 31\% alone researching a shared question ($std = 19.43$), and 38\% researching their own question ($std = 22.85$). 
The 31\% of alone-time does not fully match our impressions from the coding, but that may relate to having talked to their partner before focusing on one question (which we coded as loose collaboration). When asked about their effectiveness as a team, nine out of the 10 pairs considered that they worked together effectively or very effectively to solve the mystery.

\begin{figure}[tb]
  \centering
  \includegraphics[width=\linewidth, alt={Two slope charts comparing distances across collaboration styles, encoding the positive or negative difference between styles as the color of lines representing each participant or pair of participants. The slope chart A shows that fifteen participants stood closer to the wall during loose collaboration than during close collaboration. The slope chart B shows that eight pairs of participants stood farther away from each other during loose collaboration than during close collaboration.}]{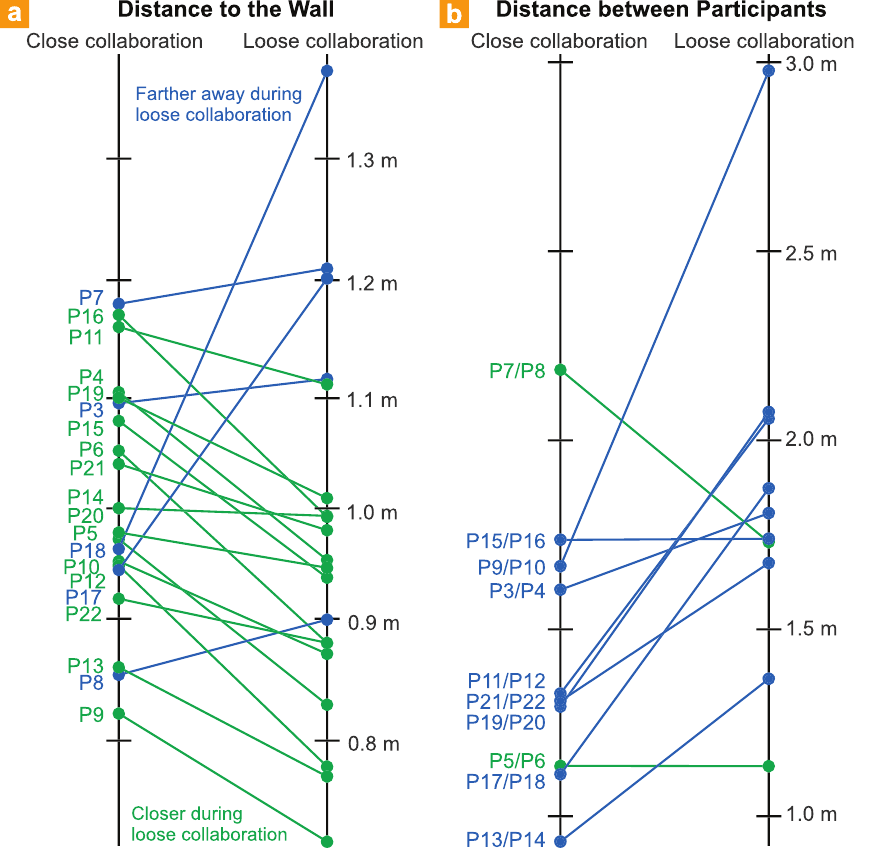}
  \caption{Slope charts comparing: (a) The distance to the wall of each participant across collaboration styles\newtext{.} (b) The distance between participants of each pair across collaboration styles.}
  \label{fig:style_and_wall_distance_merged}
\end{figure}

\Cref{fig:style_and_wall_distance_merged} (a) shows the average distance to the wall of each participant per collaboration style. Most participants stood further away from the wall while collaborating closely than while collaborating loosely. However, the difference was small, given that, on average, participants tended to stand $1.02$ m. away from the display in close collaboration, and $0.98$ m. away while collaborating loosely.
Most groups discussed and worked on the same specific problems while standing slightly further from the screen, potentially to look at the overview together (they did not say it explicitly). The most common type of close collaboration was \emph{active discussion}, which refers to when participants discussed their findings and hypotheses about the mystery they were trying to solve.
Meanwhile, the most common form of loose collaboration was working on different problems. Participants worked independently while standing closer to the screen, probably reading a document, while their partner interacted with another document related to a different problem.

We also examined how the distance between the two participants changed across collaboration styles. \Cref{fig:style_and_wall_distance_merged} (b) compares the average distance between closely and loosely coupled collaboration per pair. We found that the teams tended to be closer to each other during closely coupled collaboration than during loosely coupled collaboration. That difference was clearer and larger than in the case of the distance to the wall display. On average, participants stood at $1.45$ meters from each other while collaborating closely and at $1.90$ m.\ from each other while collaborating loosely. Therefore, they came closer or farther by half a meter. The largest difference corresponds to the pair of participants P9 and P10. Overall, they collaborated closely most of the time, but when they collaborated loosely --- mainly at the end of the session --- they tended to go to opposite ends of the display.

\subsubsection{Modality Choices During Collaboration}
We examined the relationship between input modalities and collaboration styles to find out whether participants tended to choose a specific modality while collaborating. Surprisingly, participants interacted similarly across collaboration styles. From the interactions that overlapped with a collaboration style, 53\% of the touch gestures happened during close collaboration and 47\% during loose collaboration. Regarding speech commands, it was 49\% and 51\%. Although speech interactions constituted less than 10\% of all interactions, they were evenly performed across collaboration styles, suggesting that speech commands were deemed appropriate in both loose and close collaboration.

\subsubsection{Awareness}
In the post-questionnaire, all but one participant claimed to be aware of what their partner was doing, and 18 thought their partner was aware of their actions.
Nine participants reported to have looked across the wall often to find out what their partner was doing. However, six participants from six different pairs had the impression that their partner often found important information that would have helped them, but they only learned about it later in the session.

We asked participants whether they were afraid to use speech commands as they may annoy their partner. The answers suggest that it was a concern, although not shared by the majority: Eight participants were concerned, while 11 disagreed and one was undecided. 
When on the other side participants were asked about being annoyed by the speech commands of their partner, 14 participants disagreed, and only three were actually annoyed by the commands of their partner during the study.
Furthermore, 14 participants commented positively about having learned about what their partner was doing from their speech commands, suggesting that speech interaction helped them be aware of the actions of their partner.
When we asked about the different ways they learned about what their partner was doing, participants learned by hearing the speech commands 5\% of the time on average. They knew by noticing what the partner was touching on the screen 8\% of the time.
The two most common answers were that the partner explicitly said it (24\%) and that they knew about the action because they performed it together (25\%). These answers corresponded to the groups spending most of the time in close collaboration, actively discussing.

\subsection{Interviews}
After the participants had completed their main task on the wall display, we conducted a semi-structured interview with each pair to discuss their hypotheses about the mystery, their impressions on the input modalities, and how those may have affected teamwork and awareness.

\textbf{Impressions on input modalities:}
Of the 20 participants, 16 found touch more effective than speech interaction, three found both equally effective and one person found speech more effective than touch.
Three participants mentioned that touch felt more familiar, and two referred to direct manipulation regarding using their fingers and getting immediate visual feedback (e.g., marking sentences in a document). Participants especially highlighted that moving documents and scrolling on them was faster and more precise with touch gestures. 

Although most participants found touch more suitable overall, nine mentioned specific actions for which they found speech more appropriate.  While participant P18 said \textit{``it's something about local versus global''}, P20 used the term \emph{abstract actions} to describe operations that would benefit from speech interaction, such as sorting. Two participants found speech best for actions involving multiple elements, and another one found it helpful for moving documents all the way across the screen.
Others considered speech useful for keeping an overview by interacting from a distance. P17 indicated that \textit{``if you make a search command, you can already see everything without having to move around''.} 
Unsurprisingly, accuracy was the most common issue participants mentioned concerning speech. Interestingly, seven of them mentioned that not being native English speakers played a role. At the same time, five participants explicitly separated the concept of speech interaction from the recognition accuracy they experienced. P19 expressed: 
\begin{quote}
    \textit{``I'm not that confident in the fact that it will recognize what I want (...) but my feeling is that in an ideal setup, like, maybe if the voice commands were in French, I would have used a lot more of voice commands.''} 
\end{quote}
According to P18, \textit{``maybe just the quality of speech recognition was not bad. In fact, quite good. But still (...) it's also about accents.''}
Participants also suggested additional interactions and other modalities.
To search, some wished for the virtual keyboard to move with them through proxemics and others wished for a physical keyboard to type more quickly. Two participants wished for an interaction technique to cluster documents of interest by automatically rearranging them in a screen region and moving the remaining documents to other areas.

\textbf{Teamwork:}
We asked participants whether and how the input modalities affected their teamwork. Participants discussed the advantages and disadvantages of both modalities. Some found touch helpful for highlighting sentences to silently indicate what they found relevant in a document to their partner. Yet, two participants found touch problematic because it required possibly interrupting their partner by getting close to them to interact with a document nearby. Thus, one of them preferred using speech to move documents away. However, two participants avoided using speech so as not to bother their partner, which fits the post-questionnaire answers where eight participants expressed such a concern. In contrast, four participants explained that speech was not an obstacle because they felt the difference between talking to each other and talking to the system was clear. The reasons included that speech commands required a click with the wireless mouse and each person had a microphone, so there was no need to speak loudly. P13 said \textit{``the difference between you saying a command and you addressing to me was quite clear. So it was okay if you were mumbling on your side.''} While some participants did not notice the speech commands of their partners, others said that they could not avoid listening. 
Two persons associated touch interaction with working on their own, e.g., \textit{``because the interaction makes us look at the wall and not at each other.''} 

\textbf{Awareness:} Listening to the speech commands of the other person was often associated with awareness. Five participants mentioned that listening to their partner talking to the wall helped them to have an idea of what keywords or documents the person was focusing on, which, at the same time, could be distracting: 
\begin{quote}
\textit{``The advantage of the speech interaction might be we can know what my partner is looking for because I cannot avoid to listen (...) so when he says something and I heard that, I will realize something about what he is now looking for''.}
\end{quote}

P19 mentioned that they sometimes looked at the feedback bar at the bottom of the screen to see what keywords their partner searched for.
Two participants considered touch and body movements helpful as the position of their partner would give them a hint of what documents they were working on. Eight participants expressed that talking to their partner helped them to know what they were doing.

\subsection{Design Considerations}
Based on our quantitative and qualitative findings, we draw the following design considerations for multimodal collaborative systems:
\begin{enumerate}
    \item \textbf{Speech interaction for global operations}. Speech was preferred for interactions that influenced the complete dataset, such as sorting and deselecting all documents. The interaction choices of our participants indicate that the findings of Srinivasan et al.~\cite{srinivasan21} about individuals leveraging speech commands for global actions also apply to collaborative scenarios.
    \item \textbf{Speech interaction supports closely and loosely coupled collaboration}. A likely concern about speech interaction is that it may disrupt collaborative work. However, speech was not an obstacle to collaborating closely or loosely, and participants did not walk away from each other to use speech commands. Moreover, participants used half of their speech commands during loosely coupled collaboration and the other half during close collaboration, suggesting that speech was appropriate in both cases. 
    \item \textbf{Speech interaction contributes to the awareness of the partner}. According to the post-questionnaire answers of our study, 14 of 20 participants became aware of the actions of their partner through speech commands. While some participants agreed with being concerned about annoying their partner by using speech, only three of them actually reported being annoyed by the speech commands. Therefore, including speech interaction may have a positive effect on collaborative work by increasing awareness. Given that some participants found it distracting, we should attempt to minimize distraction by considering what interactions to associate voice with, but in practice, our results indicate that likely the distraction is not high. 
    \item \textbf{Personal characteristics influence interaction choices}. From the three personality traits we examined, we found a significant relation between how agreeable a person was and the number of speech commands they used. Thus, it is important to provide alternatives so that people who do not feel comfortable with speech can still use touch. 
    Furthermore, two of our participants pointed out that being non-native speakers made them hesitant to use speech, which suggests that in a sample of native speakers only, participants may be more likely to leverage speech. Given the different factors that may influence a person's choice to use speech commands, speech interaction should not be the only input modality to support collaborative work.   
    \item \textbf{Speech interaction technologies can support probe-based research}. Our study indicated that it is possible with current technologies to create speech-based interaction prototypes for collaborative work. For the exploratory purpose of our study, we used a deep-learning solution that allowed us to recognize custom commands according to our dataset. The speech engine \cite{picovoice} processed the speech locally, which allowed us to avoid sending the audio to an external server in favor of local data privacy protection regulations. However, recognition errors still happened, so other solutions should be tested. Using a wireless device to activate the speech recognition was well received by the participants. 
\end{enumerate}

\section{Discussion and Future Work}
Speech interaction is considered a promising input modality because it allows people to express what they wish for, instead of taking the time to translate it into code, formulas, or widget input, and has the potential to make interactive visualizations more accessible \cite{hoque18}. However, it often suffers from a lack of trust due to recognition errors \cite{baughan23} and past work has shown that it can discourage parallel work \cite{tse08}. Our findings show that speech commands can support distant interaction during collaboration without major issues but tend to be used only for specific actions and still suffer from technological challenges. 

We used up-to-date tools powered with deep learning to enable speech interaction, but recognition errors were still a common problem, and two participants seemed to give up on speech after some errors. For instance, P17 said in the interview: \textit{``I believe I stopped using speech because I couldn't get it to understand''}.
Compared with the strong preference for speech command interaction on large displays in a previous study \cite{molina24}, it seems that the recognition accuracy of current solutions still needs to catch up to support the use of speech commands as often as people actually may want to use them. 
As P8 indicated, \textit{``for example, [commercial voice assistant], they worked on that many years and they're, they still have problems''}.
In addition, people may take speech recognition issues personally. In our study, a common concern among participants was that a foreign accent may be the cause of the errors. That may be plausible because the only native speaker had only four errors during the whole session (the maximum was 36), but a more diverse sample is needed to assess the effect.
Overall, there were 150 speech errors. In case of an error, participants switched to touch in 35 cases. In 61 cases, the person insisted on using speech until being successful. In 14 cases, the person tried using speech up to four more times before switching to touch.
Another unexpected challenge we faced during the system implementation was that the speech engine did not recognize multiple names as English words. For example, the town name in the original\st{ VAST} dataset was Alderwood, but the engine did not recognize it. Thus, we had to modify some terms from the dataset.

\subsection{Facilitating Specific Actions}
Despite the challenges, participants clearly favored speech interaction for global operations such as sorting and \textit{deselect all}. Moreover, search commands via speech were common and were even part of the conversation within some teams.
P19 indicated that they started working on the task by discussing what to focus on and used speech: \textit{``because we were together, it was like implicit that we were talking to the wall (...) then we went apart and it was not that clear and we focused on touch''}.
The sensemaking task promoted collaborative work but relied heavily on browsing and reading multiple documents. Some participants argued that using a speech command to open or move a document was cumbersome because it required first identifying the document by its number. That suggests that speech interaction is more suitable for group or set-related actions, such as in prior work about data exploration by individuals, e.g., encoding an attribute with color \cite{srinivasan18}. Some participants indicated that speech commands helped them interact with multiple documents that were either far from the person or far from each other on the screen.
In the preceding elicitation study \cite{molina24}, participants wished to filter data, change the visual encoding, toggle brushing \& linking, and merge and split multiple views via speech. These operations involved the whole dataset or all views, corresponding to global operations and reaching far-away elements.
In the context of visual analytics, advanced operations such as invoking automated analysis methods on the dataset may be a good fit for speech interaction.

While most participants were not disturbed by the speech commands of their partner, these commands represent less than 10\% of the logged interactions. To put that into context, 15\% of the interactions correspond to participants moving documents via touch.
Maybe there was little conflict because the speech commands were so few in contrast to touch gestures, and, therefore, it was unlikely that both persons would try to talk to the wall simultaneously. Global operations, such as sorting, may indeed happen less often than local operations. Thus, speech may be a modality used less often than others that suit more common actions, e.g., touch for scrolling.
A follow-up question would be whether there is a collaborative scenario where speech commands would prevail. Then, the likelihood of collaboration conflicts may rise.

\subsection{Supporting Co-located Collaboration}
Regarding the characteristics of collaborative work, our study focused on examining how speech interaction may affect group communication, the distance between team members, and awareness. In contrast to the findings of Tse et al.~\cite{tse08} on tabletops, our participants did not find speech interaction problematic for working in parallel. One reason may be the larger space to interact in front of the wall display. Another one may be the speaking volume, given that some participants referred to not having to speak loudly with the wireless microphone. 
Given the concern of some participants about distracting others through speech interaction, one may consider including dedicated awareness features. For example, Cernea et al.~\cite{cernea15} proposed visualizing emotions through \textit{emotion-prints} to increase emotional awareness among collaborators. Such techniques may help assess whether specific actions negatively affect any collaborator.
There are also other social aspects relevant to collaboration that may be affected by speech interaction. Future work should investigate how speech may affect privacy and territoriality \cite{azad12}. While our participants did not report any inconveniences, working separately and keeping interactions private may be more relevant in other collaborative scenarios. For instance, P3 said that touch may be more appropriate for \textit{``privacy, maybe. Not particularly in this context, but (...)  if you work on sensitive data.''}
Participants appreciated that the speech recognition had to be manually activated to avoid unintended interactions with the system while talking to each other.
Regarding territoriality, previous work found that the size of a personal space changes across cultures \cite{hall66}. Therefore, the distance between our participants may have been influenced by their cultural background. However, we did not control cultural backgrounds and can, therefore, not provide supporting data. Moreover, we had participants with similar cultural backgrounds and different behaviors. Thus, personality differences may be more important. Still, speech allows distant interaction with elements in the personal space of others. Future work could investigate the relationship between speech interaction and social norms.

While the interaction techniques that our system supported used one modality at a time, future work should consider the combination of two modalities in one action. Our participants suggested combining touch and speech for some actions, such as tapping on a document and then saying \textit{``Tag this with police''} or tapping on an empty screen space followed by \textit{``Bring [document ID] over here''} to move documents. These interactions could also be collaborative, i.e., performed jointly by two persons. However, such interactions require good coordination and may come with technical challenges, such as distinguishing which action comes from which person. Moreover, the chance of having overlapping voices increases with the number of collaborators, which may make speech recognition more challenging.

Finally, we did not focus on performance, as our study was rather exploratory. An important future research direction would be to investigate how the use of speech commands may influence performance in the context of collaborative work. In the interview, participants often favored touch over speech due to speed and accuracy. However, the participant with the highest number of successful speech commands argued that getting used to speech helped to use it more. Familiarity may thus play an important role in ensuring performance via speech.

Which other visual analytics tasks would benefit from leveraging speech? Future work should systematically explore interaction taxonomies to determine when is speech suitable.
Moreover, with larger groups, it would be possible for two persons to talk while a third one interacts via speech. Would speech become a significant obstacle to communication then?
With our findings, we hope to encourage researchers and practitioners to consider speech interaction for interactive systems not only for individuals but also for collaborative scenarios.

\section{Conclusion}
We conducted an in-depth exploratory study on combining touch gestures and speech commands for collaborative sensemaking to extend our understanding of whether and how speech interaction can be leveraged for co-located collaborative work. Our findings provide evidence that speech commands are a viable option to support global operations and distant interaction from the wall display. 
The differences between modality preferences and actual use suggest that in concordance with previous work \cite{molina24}, people are interested in using speech interaction, but other aspects, such as speech recognition errors and the lack of familiarity, may influence their interaction choices on implemented systems. Furthermore, we found the tendency to use speech commands correlated with the personality trait of agreeableness. Future work can investigate whether the choice to leverage other input modalities may be influenced by personal characteristics.
Regarding the communication between team members, we did not find enough evidence that speech interaction was an obstacle to collaboration. On the contrary, 14 of 20 participants reported that speech commands contributed to becoming aware of their partner's actions. Based on our findings, we introduced design considerations for incorporating speech commands to collaborative systems and highlight opportunities for future research.
Our collaborative visual analytics prototype, \oursystem, which combines touch and speech interaction for collaborative sensemaking, is openly available together with our supplemental material.

\section*{Supplemental Materials}
\label{sec:supplemental_materials}

All supplemental materials are available on OSF at \url{https://osf.io/8gpv2}, released under a CC BY 4.0 license.
In particular, they include (1) the questionnaires, (2) the answers to the questionnaires, and (3) the interaction logs.
The open-source code of the system is available at \href{https://gitlab.inria.fr/aviz/TouchTalkInteractive}{https://gitlab.inria.fr/aviz/TouchTalkInteractive}.

\acknowledgments{%
  This work was supported by a fellowship of the German Academic Exchange Service (DAAD) as well as by the French National Research Agency (ANR) under the Investments for the Future program (PIA) grant ANR-21-ESRE-0030 (CONTINUUM). An earlier version of this paper is part of the dissertation of the first author \cite{molina24b}.%
}

\bibliographystyle{abbrv-doi-hyperref}

\bibliography{template}

\newpage
\appendix 

\end{document}